\documentclass[useAMS,usegraphicx,usenatbib,usedcolumn,preprint]{aastex}           
\usepackage{spr-astr-addons}
\usepackage{graphicx,amssymb,amsmath,times}
\usepackage{epsfig,epstopdf}
\usepackage[colorlinks=true,pdfstartview=FitV,linkcolor=blue,citecolor=blue,urlcolor=blue]{hyperref}  
\usepackage{amsmath} % align equations - arrays

\def\Journal#1#2#3#4{{#4}, {#1}, {#2}, #3} 

\newcommand{\etal}{et al.}

\newcommand{\AMS}{\textsf{AMS}} 
\newcommand{\ApJ}{ApJ}
\newcommand{\AeA}{A\&A}
\newcommand{\ApP}{APh}
\newcommand{\PRD}{PRD}
\newcommand{\PRC}{PRC}

\newcommand{\MNRAS}{MNRAS}
\newcommand{\NIM}{NIM}

\newcommand{\He}{\textsf{He}}
\newcommand{\Hyd}{\textsf{H}}

\newcommand{\Hu}{\ensuremath{^1}\textsf{H}}   % $^{1}$H
\newcommand{\Hd}{\ensuremath{^2}\textsf{H}}   % $^{2}$H
\newcommand{\Ht}{\ensuremath{^3}\textsf{H}}   % $^{3}$H
\newcommand{\Het}{\ensuremath{^3}\textsf{He}} % $^{3}$He
\newcommand{\Heq}{\ensuremath{^4}\textsf{He}} % $^{4}$He

\newcommand{\HdHet}{\ensuremath{^2}\textsf{H}/\ensuremath{^3}\textsf{He}}   %$^{2}$H/$^{3}$He
\newcommand{\HdHeq}{\ensuremath{^2}\textsf{H}/\ensuremath{^4}\textsf{He}}   % $^{2}$H/$^{4}$He
\newcommand{\HetHeq}{\ensuremath{^3}\textsf{He}/\ensuremath{^4}\textsf{He}} %$^{3}$He/$^{4}$He

\newcommand{\LiRatio}{\ensuremath{^6}\textsf{Li}/\ensuremath{^7}\textsf{Li}}  %$^{6}$Li/$^{7}$Li
\newcommand{\BRatio}{\ensuremath{^{10}}\textsf{B}/\ensuremath{^{11}}\textsf{B}} %$^{10}$B/$^{11}$B

\newcommand{\BC}{\textsf{B}/\textsf{C}}  %B/C
\newcommand{\LiBe}{\textsf{Li}/\textsf{Be}}  %Li/Be

\newcommand{\Bes}{\ensuremath{^7}\textsf{Be}}   %$^{7}$Be
\newcommand{\Ben}{\ensuremath{^9}\textsf{Be}}   %$^{9}$Be
\newcommand{\Bed}{\ensuremath{^{10}}\textsf{Be}} %$^{10}$Be

\begin{document}

%%%%%%%%%%%%%%%%%%%%%%%%%%%%%%%%%%%%%%%%%%%%%%%%%%%%%%%%%%%%%%%
\title{Propagation of \Hyd{} and \He{} Cosmic Ray Isotopes in the Galaxy: \\ Astrophysical and Nuclear Uncertainties}
\shorttitle{\Hyd{} and \He{} Isotopes in CRs: Astrophysical and Nuclear Uncertainties}                   %%
%%%%%%%%%%%%%%%%%%%%%%%%%%%%%%%%%%%%%%%%%%%%%%%%%%%%%%%%%%%%%%%

\author{Nicola Tomassetti}
\affil{\it INFN - Sezione di Perugia, 06122 Perugia, Italy. \\E-mail: nicola.tomassetti@pg.infn.it}
\shortauthors{Nicola Tomassetti\altaffilmark{1}}

%%%%%%%%%%%%%%%%%%%%
\begin{abstract} %%%
%%%%%%%%%%%%%%%%%%%%

Observations of light isotopes in cosmic rays provide valuable information on their origin and propagation in the Galaxy. 
Using the data collected by the \AMS-01 experiment in the range $\sim$0.2\,\--\,1.5\,GeV\,nucleon$^{-1}$, 
we compare the measurements on \Hu, \Hd, \Het, and \Heq{} with calculations for interstellar propagation and solar modulation. 
These data are described well by a diffusive-reacceleration model with parameters that match the \BC{} ratio data,
indicating that \He{} and heavier nuclei such as \textsf{C-N-O} experience similar propagation histories. 
Close comparisons are made within the astrophysical constraints provided by the \BC{} ratio data and within the nuclear 
uncertainties arising from errors in the production cross section data.
The astrophysical uncertainties are expected to be dramatically reduced by the data upcoming from \AMS-02,
so that the nuclear uncertainties will likely represent the most serious limitation on the reliability of the model predictions.
On the other hand, we find that secondary-to-secondary ratios such as \HdHet{}, \LiRatio{} or \BRatio{}  are barely sensitive to the
key propagation parameters and can represent a useful diagnostic test for the consistency of the calculations. 
\end{abstract}

%%%%%%%%%%%%%%%%%%%%%%%%%%%%%
\section{Introduction}    %%%
\label{Sec::Introduction} %%%
%%%%%%%%%%%%%%%%%%%%%%%%%%%%%

Secondary Cosmic Ray (CR) isotopes such as \Hd{}, \Het{} and \textsf{Li-Be-B} are believed to be produced as a results of  
of nuclear interactions primary CRs such as \Hu{}, \Heq{} or \textsf{C-N-O} with the gas nuclei of the interstellar medium (ISM). 
The secondary CR abundances depend on the intensity of their progenitors nuclei, their production rate 
and their transport in the turbulent magnetic fields \citep{Strong2007}.
Secondary to primary ratios such as \HdHeq{}, \HetHeq{} or \BC{}
be used to study the CR propagation processes in the Galaxy. The \BC{} ratio is widely 
used to determine the key parameters of propagation models. In fact the \BC{} ratio is measured by several experiments 
between $\sim$\,100\,MeV and $\sim$\,1\,TeV of kinetic energy per nucleon. 
The CR propation physics is also connected with the indirect search of dark matter particles. 
In this context the CR propagation models, once tuned to agree with the \BC{} ratio, are used to compute the secondary production for other 
rare species such as $\bar{p}$ or $\bar{d}$, that provides the \textit{astrophysical background} for the 
search of new physics signals \citep{Donato2008,Evoli2011,Salati2010}. 
Clearly, understanding the CR propagation processes is crucial for modeling both the CR signal and the background.
Furthermore, these studies assume that all the CR species experience the same propagation effects in their 
journey thourghout the ISM \citep{Putze2010,Trotta2011}. It is therefore important to test the CR 
propagation with nuclei of different mass-to-charge ratios. This issue of the \textit{universality} of CR 
propagation histories was also studied in \citet{Webber1997} and, recently, in \citet{Coste2011}. 

In this work we use the recent \AMS-01 observations for the \HdHeq{} and \HetHeq{} ratios and compare 
them with the expected ratios based on interstellar and heliospheric propagation calculations. The aim of this work 
is to determine wether the \AMS-01 observations are consistent with the propagation calculations derived from 
heavier nuclei (mainly from \BC{} data). This consistency is inspected within two classes of model uncertainties: 
the \textit{astrophysical uncertainties}, which are related to the knowledge of the CR transport parameters given 
by the \BC{} ratio, and the \textit{nuclear uncertainties}, which arise from the \Hd{} and \Het{} production cross sections.

%%%%%%%%%%%%%%%%%%%%%%%%%%%%%%%%
\section{Observations}       %%%
\label{Sec::AMSObservations} %%%
%%%%%%%%%%%%%%%%%%%%%%%%%%%%%%%%

The \AMS-01 experiment operated successfully in the STS-91 mission on board the space shuttle \textit{Discovery}. 
The spectrometer was composed of a cylindrical permanent magnet, a silicon micro-strip tracker, time-of-flight 
scintillator planes, an aerogel \v{C}erenkov counter and anti-coincidence counters. The performance of \AMS-01 
is described elsewhere~\citep{AMS01Report2002}. Data collection started on 1998 June 3 and lasted 10 days. 
\AMS-01 observed cosmic rays at an altitude of $\sim\,$380 km during a period, 1998 June, of relatively quiet 
solar activity. Results on isotopic spectra have been recently published in \citep{AMS01Isotopes2011} with
the ratios \HdHeq{}, \HetHeq{}, \LiRatio{}, \Bes{}/(\Ben{}+\Bed{}) and \BRatio{} 
in the range $\sim 0.2\--1.5$ GeV of kinetic energy per nucleon. 
\begin{figure*}[!htb]
\begin{center}
\epsscale{1.80}
\plotone{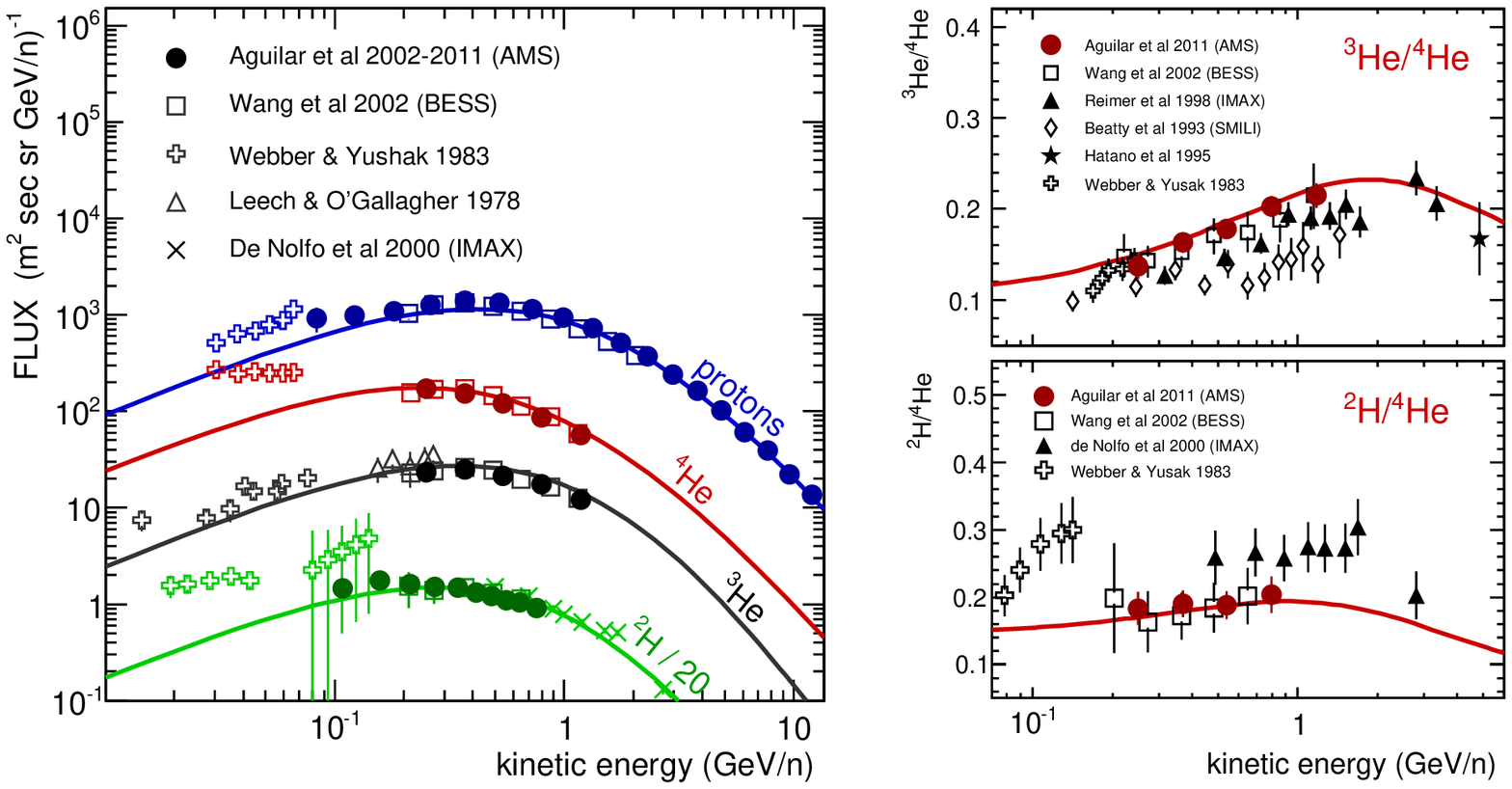}
\figcaption{ 
  Left: energy spectra of CR proton, deuteron (divided by 20), \Het{} and \Heq{}. Right: isotopic ratios \HdHeq{} and \HetHeq{}.
  Calculations are compared with the data from 
  \AMS-01 \citep{AMS01Report2002,AMS01Isotopes2011},
  IMAX \citep{Reimer1998,DeNolfo2000}, SMILI \citep{Ahlen2000}, 
  BESS \citep{Wang2002}, \citet{Hatano1995}, \citet{Leech1978}, \citet{WebberYushak1983}. 
  \label{Fig::ccIsotopeFluxes}
}
\end{center}
\end{figure*}
Fig.\,\ref{Fig::ccIsotopeFluxes} shows the \AMS-01 energy spectra of proton, deuteron, helium isotopes, and the 
ratios \HdHeq{} and \HetHeq{}. The other data come from balloon borne experiments IMAX 
\citep{Reimer1998,DeNolfo2000}, SMILI \citep{Ahlen2000}, BESS \citep{Wang2002}, \citet{Hatano1995}, \citet{Leech1978} 
\citet{WebberYushak1983}.

The \AMS-01 observations are made in a period, 1998 June, of relatively quiet solar activity, and the particle 
recorded are free from any atmospheric induced background. Furthermore, the \AMS-01 material thickness between 
the top of the payload and the active detector amounted to $\sim$\,5\,g\,cm$^{-2}$ which is considerably less 
than that of previous balloon borne experiments ($\sim$\,9--20\,g\,cm$^{-2}$ of top-of-instrument material plus 
$\sim$\,5\,g\,cm$^{-2}$ of residual atmosphere). 
Also important, for the aims of this work, is to realize that high precision data are currently flowing in from 
two active projects PAMELA and \AMS-02, both operating in space. In particular, the data forthcoming by
%\AMS-02 \footnote{\url{www.ams02.org/ams-and-iss/data-acquisition}}
%{http://www.ams02.org}}
are expected to provide a dramatic improvement in our understanding of the CR transport processes and interactions \citep{NTFD2012,Coste2011,Oliva2008}.

%%%%%%%%%%%%%%%%%%%%%%%%%%%%%%%%%%%%%%%%%%%%
\section{CR Transport and Interactions}  %%%
\label{Sec::CRPropagation}               %%%
%%%%%%%%%%%%%%%%%%%%%%%%%%%%%%%%%%%%%%%%%%%%

Galactic CR nuclei are believed to be accelerated by particle diffuse shock acceleration mechanisms 
occurring in galactic sites such as supernova remnants (SNRs). Their propagation in the ISM is dominated 
by particle transport in the turbulent magnetic field and interactions with the matter, that is 
generally described by a diffusion-transport equation including source distribution functions, 
magnetic diffusion, energy losses, hadronic interactions, decays, diffusive reacceleration and convective transport
(the latter is not considered in this work).
Models of CR propagation in the Galaxy employ fully analytical \citep{Thoudam2008,NT2012},
semi-analytical \citep{Jones2001,Putze2010}, or fully numerical calculation frameworks \citep{DiBernardo2010,Strong2007}.
The present work relies on the \textit{diffusive-reacceleration} model implemented with \texttt{GALPROP-v50.1p}\footnote{\url{http://galprop.stanford.edu}}, 
which numerically solves the cosmic ray propagation equation for a cylindrical diffusive region with 
a realistic interestellar gas distribution and source distribution.
\begin{figure*}[!htb]
\begin{center}
\epsscale{1.80}
\plotone{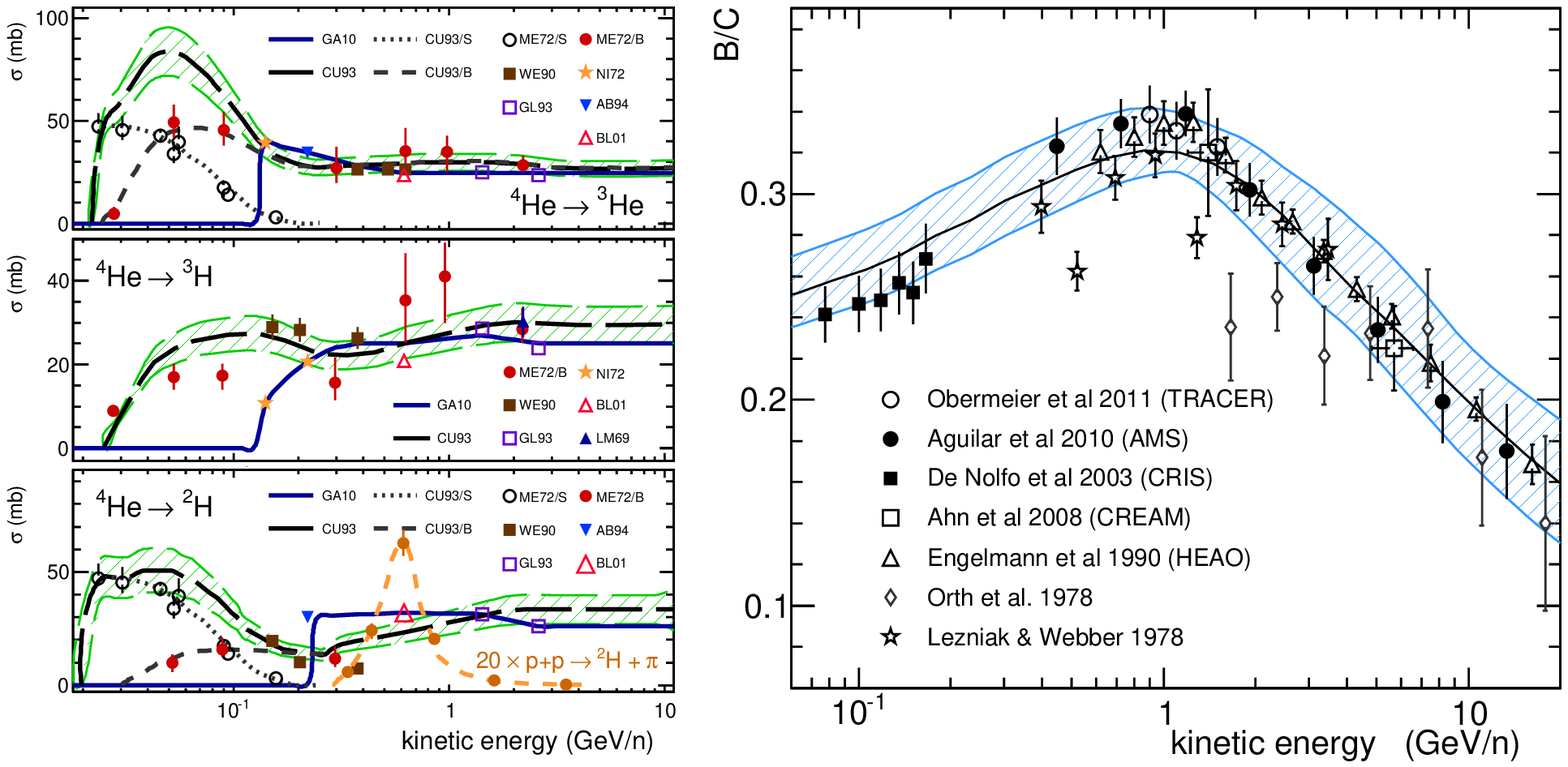}
\figcaption{ 
  Left: Cross section parametrizations \citep{Cucinotta1993} for the channels \Heq{}$\rightarrow$\Het{},
  \Heq{}$\rightarrow$\Ht{}, \Heq{}$\rightarrow$\Hd{}
  and $p+p \rightarrow \pi + ^{2}$H (multiplied by 20).
  The data are encoded as ME72: \citet{Meyer1972}, WE90: \citet{Webber1990}, NI72: \citet{Nicholls1972},
  GL93: \citet{Glagolev1993}, AB94: \citet{Abdullin1994}, BL01: \citet{Blinov2001}, LM49: \citet{Lebowitz1969},
  Right: \BC{} ratio from the CR propagation model of Table\,\ref{Tab::ModelParameters} and 1--$\sigma$
  uncertainty band. Data are from TRACER \citep{Obermeier2011}, \AMS-01 \citep{AMS01Nuclei2010},
  CRIS \citep{DeNolfo2003}, CREAM \citep{Ahn2008}, HEAO \citep{Engelmann1990}, \citet{Orth1978}, and \citet{Lezniak1978}.
  \label{Fig::ccHeliumDeuteronXS}
}
\end{center}
\end{figure*}

%%%%%%%%%%%%%%%%%%%%%%%%%%%%%%%%%%%%%%%%%%%%%%%%%
\subsection{Diffusive\--Reacceleration Model} %%%
\label{Sec::DiffusionReacceleration}          %%%
%%%%%%%%%%%%%%%%%%%%%%%%%%%%%%%%%%%%%%%%%%%%%%%%%

The propagation equation of the diffusive-reacceleration model for a CR species $j$ is given by: %~\cite{Strong2007}
\begin{align}\label{mastereq}
  \frac{\partial \mathcal{N}_{j}}{\partial t} = & q^{tot}_{j}+\vec{\nabla}\cdot\left(D\vec{\nabla}\mathcal{N}_j\right) - {\mathcal{N}_j}{\Gamma^{tot}_{j}} \nonumber \\ 
+& \frac{\partial}{\partial p} p^2 D_{pp} \frac{\partial}{\partial p} p^2 \mathcal{N}_j - \frac{\partial}{\partial p}\left(\dot{p}_{j} \mathcal{N}_{j}\right) % \nonumber \\
\end{align}
where $\mathcal{N}_{j}=dN_{j}/dVdp$ is the CR density of the species $i$ per unit of total momentum $p$. 
%%%%%% PROPAGATION PARAMETER SET %%%%%%%%%%%%%%%%%%%%%%%%
\begin{table}
\small
\caption{Propagation parameter set.\label{Tab::ModelParameters}}
\begin{tabular}{llc}
\tableline
Parameter & Name & Value \\
\tableline
Injection, break value & $R_{B}$ [GV]     & 9 \\
Injection, index below $R_{B}$ &    $\nu_{1}$ & 1.82 \\  
Injection, index above $R_{B}$ &    $\nu_{2}$ & 2.36 \\

Diffusion, magnitude    &  $D_{0}$ [cm$^2$ s$^{-1}$]  & $5.75\cdot10^{28}$ \\
Diffusion, index        &   $\delta$                 & 0.34 \\
Diffusion, ref. rigidity & $R_{0}$ [GV]              & 4 \\

Reacceleration, Alfv\'en speed & $v_{A}$ [km s$^{-1}$]  & 36  \\

Galactic halo, radius  &   $\mathcal{R}$ [kpc] & 20  \\
Galactic halo, height  &   $z_{h}$ [kpc]        & 4   \\

Solar modulation parameter &  $\phi$ [MV]   & 500  \\
\tableline
\end{tabular}
\end{table}
%%%%%%%%%%%%%%%%%%%%%%%%%%%%%%%%%%%%%%%%%%%%%%%%%%%%%%%%%%%%%%%%%%%%%%%%%%%
\begin{figure*}[!htb]
\begin{center}
\epsscale{1.80}
\plotone{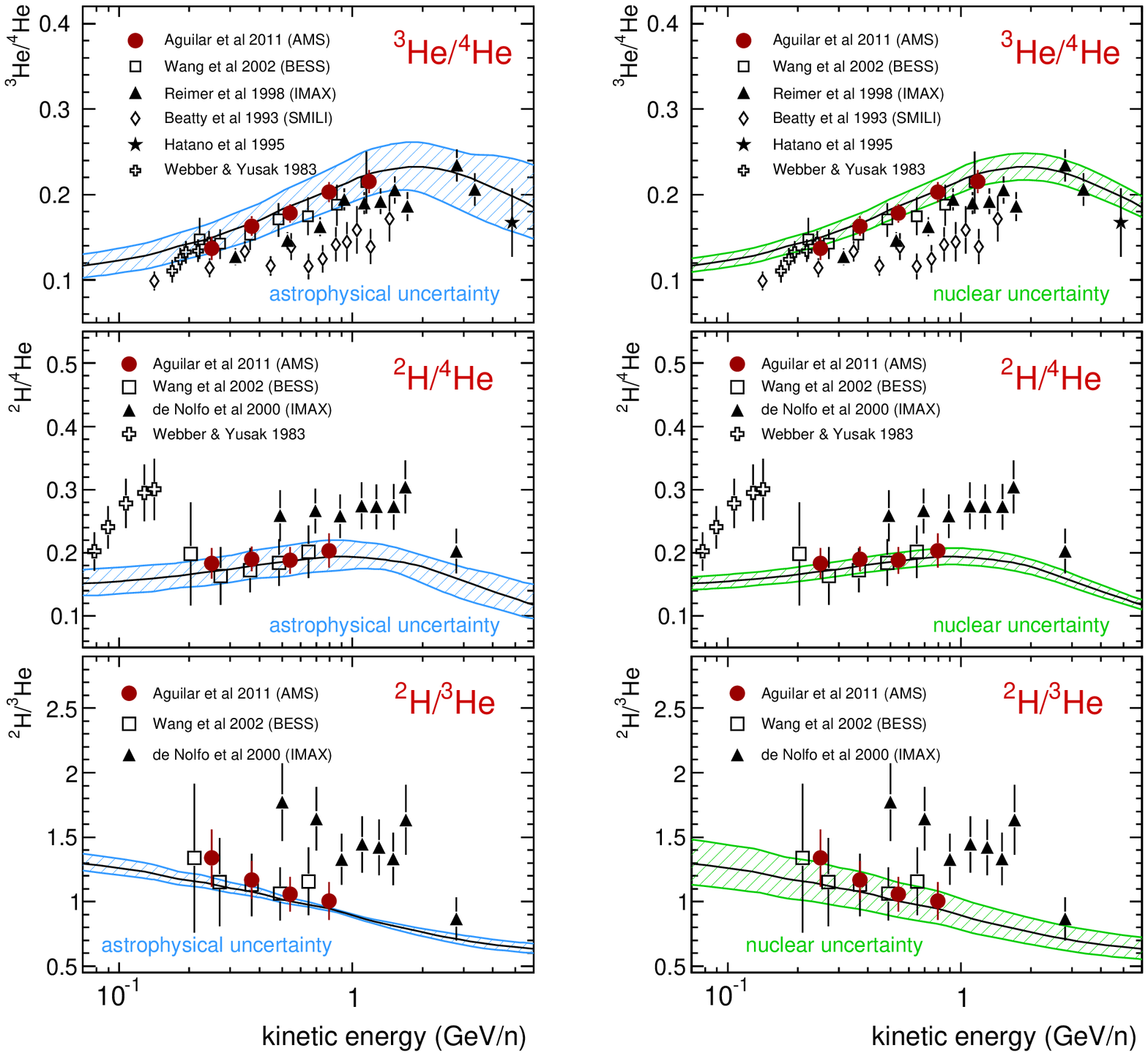}
\figcaption{ 
  Astrophysical (left) and nuclear (right) uncertainty bands for the predicted ratios \HdHeq{}, \HetHeq{} and \HdHet{}
  in comparison with the \AMS-01 data. Other data are from IMAX \citep{Reimer1998,DeNolfo2000}, SMILI \citep{Ahlen2000}, BESS \citep{Wang2002}
  \citet{Hatano1995}, and \citet{WebberYushak1983}.
  \label{Fig::ccIsotopeRatios2X2}
}
\end{center}
\end{figure*}

The source term, $q^{\rm tot}_{j}=q^{\rm pri}_{j} + q^{\rm sec}_{j}$, includes the primary acceleration spectrum 
(\textit{e.g.} from SNRs) and the term arising from the secondary production in the ISM or decays. 
The primary spectrum is $q_{j}^{\rm pri} = q_{0}\left(R/R_{B}\right)^{-\nu}$, is normalized to 
the abundances, $q_{j}^{0}$, at the reference rigidity $R_{B}$. The injection spectral indices are
the same for all the primary elements. The source spatial distribution in the galactic disc is extracted 
from SNR observations. The secondary production term, 
$q_{j}^{sec} = \sum_{k} \mathcal{N}_{k} \Gamma_{k\rightarrow j} $, describe the products of decay and spallation 
of heavier CR progenitors with numer density $\mathcal{N}_{k}$. For collisions with the interstellar 
gas:
\begin{equation}
  \Gamma_{k\rightarrow j} =  \beta_{k}c \sum_{\rm ism} \int_{0}^{\infty} n_{\rm ism} \sigma_{k\rightarrow j}^{\rm ism}(E,E') dE'\,, 
\end{equation} 
where $n_{\rm ism}$ are the number densities of the ISM nuclei, $n_{H}\approx 0.9\,cm^{-3}$ and $n_{He}\approx 0.1\,cm^{-3}$,  
and $\sigma_{k \rightarrow j}^{\rm ism}$ are the fragmentation cross sections for the production of a $j$-type species at energy 
$E$ from a $k$-type progenitor of energy $E'$ in H or He targets. 
$\Gamma^{tot}_{j}=\beta_{j}c \left( n_{H}\sigma_{j,H}^{tot} + n_{He}\sigma_{j,He}^{tot} \right) + \frac{1}{\gamma_{j}\tau_{j}}$ 
is the total destruction rate for inelastic collisions (cross section $\sigma^{tot}$) and/or 
decay for unstable particles (lifetime $\tau$). The spatial diffusion coefficient $D= D(\textbf{r},p)$ is taken as 
spatially homogeneus and rigidity dependent as $D(R)=\beta D_{0}\left(R/R_{0}\right)^{\delta}$, where $R=pc/Ze$ is the 
rigidity, $D_{0}$ fixes the normalization at the reference rigidity $R_{0}$, and the parameter $\delta$ 
specifies its rigidity dependence. Diffusive reacceleration is described as diffusion process acting in momentum 
space. It is determined by the coefficient $D_{pp}$ for the momentum space diffusion:
\begin{equation}
D_{pp} = \frac{4 p^{2} v_{A}}{3 \delta \left( 4 - \delta^{2} \right)\left(4-\delta \right)D }
\end{equation}
where $v_{A}$ is the Alfv\'en speed of plasma waves moving through the ISM.
The last term describes Coulomb and ionization losses by means of the momentum loss rate $\dot{p}_{j}$.

\texttt{GALPROP} solves the steady-stady equation $\partial \mathcal{N}_{j} / \partial t = 0$ for 
a cylindrical diffusion region of radius $r_{max}$ and half-thickness $L$ with boundary conditions
$\mathcal{N}_{j}$($r$=$r_{max}$)=0 and $\mathcal{N}_{j}$($z$=$\pm$$L$)=0. 
The local interstellar spectrum (LIS) is then computed for each species at the solar system coordinates ${r_{\odot}}=8.5\,kpc$ and $z=0$:
\begin{equation}\label{Eq::PhiLIS}
\Phi^{\rm LIS}_{j}(E) = \frac{c A}{4\pi}\mathcal{N}_{j}(r_{\odot},p) \,,
\end{equation}
where $A$ is the mass number and the flux $\Phi^{\rm LIS}$ is given in units of kinetic energy per nucleon $E$.
For the descritpion we adopt the ``conventional model'' which finely reproduces the CR elemental 
fluxes at intermediate energies of $\sim$\,100\,MeV -- 100\,GeV per nucleon.

%%%%%%%%%%%%%%%%%%%%%%%%%%%%%%%%%%%%%%%%%%
\subsection{Heliospheric Propagation}  %%%
\label{Sec::SolarModulation}           %%%
%%%%%%%%%%%%%%%%%%%%%%%%%%%%%%%%%%%%%%%%%%

CRs in the solar neighbourhood undergo convection, diffusion and energy changes 
as results of the expansion of the solar wind. To describe the solar modulation effect, 
we adopt the so-called \textit{force-field approximation} that arise from the a 
spherically symmetric description of the heliosphere \citep{Gleeson1968}. 
The correspondence between the (modulated) top-of-atmosphere spectrum, $\Phi^{\rm TOA}$, 
and the (unmodulated) LIS spectrum of \ref{Eq::PhiLIS}, $\Phi^{\rm LIS}$, is expressed by 
the effective parameter $\phi$ (GV) through:
\begin{equation}\label{Eq::SolarMod}
  \Phi^{\rm IS} \left(E^{\rm IS} \right) = \left( \frac{p^{\rm IS}}{p}\right)^{2} \Phi^{\rm TOA} \left(E^{\rm TOA}\right) \,,
\end{equation}
where the LIS and TOA energies per nucleon are related by $E^{\rm IS} = E^{\rm TOA} + \frac{Z}{A}\phi$.
The main parameters of the model are listed in Table\,\ref{Tab::ModelParameters}. 
The remaining specifications are as in the file \texttt{galdef\_50p\_599278} provided with the package.

%%%%%%%%%%%%%%%%%%%%%%%%%%%%%%%%%%%%%%%%%%%%%%
\subsection{Fragmentation Cross Sections}  %%%
\label{Sec::FragmentationCrossSections}    %%%
%%%%%%%%%%%%%%%%%%%%%%%%%%%%%%%%%%%%%%%%%%%%%%

To compute the secondary nuclei production rate from the disintegration of the heavier CR nuclei, a large amount 
of cross section estimates is required. The accuracy of the calculated secondary spectra therefore depends on 
the reliability of the production and destruction cross sections employed.

The production of \Hd{} and \Het{} isotopes is mainly due to collision of \Heq{} nuclei. The \Het{} isotopes 
are also produced via decay of trithium (\Ht{}\,$\rightarrow$\,\Het{}) which, in turn, is predominantly created 
by \Heq{} spallation. The most relevant \textit{projectile}~$\rightarrow$~\textit{fragment} processes for the 
\Het{} abundance are \Heq{}\,$\rightarrow$\,\Het{} and \Heq{}\,$\rightarrow$\,\Ht{}. The main deuteron 
production channel is \Heq{}\,$\rightarrow$\,\Hd{}. Low energy deuterons are also created by the fusion reaction 
$p+p \rightarrow \pi + ^{2}$H acting between $\sim$\,300 and $\sim$\,900\,MeV of the proton energy \citep{Meyer1972}.
Although the $p$-$p$ fusion cross section is very small ($\sigma\sim$\,3\,mb), this reaction 
contributes appreciably to the \Hd{} abundance because of the large CR proton flux.
Spallation of heavier nuclei (C, O, Fe) give a minor contribution, roughly $\lesssim$\,10\,\% of their fractional 
abundance. For all these channels the total inclusive reaction can be realized in a number of possible final states. 
We employ the phenomenological parametrization of \citet{Cucinotta1993}. These parametrizations are shown in 
Fig.~\ref{Fig::ccHeliumDeuteronXS} together with the accelerator data.  For \Hd{} and \Het{}, the partial 
contributions of break-up (B) and stripping (S) reactions are shown separately.

For all these processes we have assumed that the fragment is ejected with the same the kinetic energy per nucleon $E$ of 
the projectile, $E'$. This \textit{straight-ahead} approximation, expressed by $\sigma(E,E')\approx\sigma(E)\delta(E-E')$,
has been validated within some percent of accuracy for reactions involving $Z>2$ nuclei \citep{Kneller2003} and for 
lighter species ($Z<3$) \citep{Cucinotta1993}. For the $p$-$p$ fusion channel, the kinetic energy per nucleon is not 
conserved due to the kinematic of the reaction: the energy of ejected deuterons is sistematically lower than that of the 
proton of a factor $\sim$\,4 on average. Using a \textit{straight-ahead} fashion, we write 
$\sigma(E,E')\approx\sigma(E)\delta(E-\xi E')$, where $\xi\cong$\,4 is the average inelasticity of the \Hd{}.
The $p$-$p$ fusion contributes to the \Hd{} flux at energies below $\sim$\,250\,MeV\,nucleon$^{-1}$. 
For nuclear reactions involving heavier ($Z>2$) nuclei, we use the default cross section parametrization of 
\texttt{GALPROP}. To account for CR collisions with the interstellar helium ($\sim$\,10\% of the ISM) the parametrization 
of \citet{Ferrando1988} is used. Calculation of CR spectra and ratios \HdHeq{} and \HetHeq{} are shown in 
Fig.\,\ref{Fig::ccIsotopeFluxes} for the modulation intensity of $\phi$\,=\,500\,MV.

%%%%%%%%%%%%%%%%%%%%%%%%%%%%%%%%%%%%
\section{Model Uncertainties}    %%%
\label{Sec::ModelUncertainties}  %%%
%%%%%%%%%%%%%%%%%%%%%%%%%%%%%%%%%%%%

We consider two classes of uncertainty in the model estimates. The \textit{astrophysical uncertainties} are 
those associated with the transport parameters constrained by the \BC{} ratio. The relevant parameters for the 
secondary productions are $\delta$, $v_{A}$ and the ratio $D_{0}/L$. We perform a grid scan in the parameter 
space $\{\delta, v_{A}, D_{0}/L\}$ by running \texttt{GALPROP} multiple times. The other parameters, \textit{e.g.} 
the source parameters and the modulation potential, are kept fixed. For each parameter configuration, we select 
the \BC{}-compatible models within one sigma of uncertainty in the $\chi^{2}$ statistics. We use \BC{} data from 
HEAO \citep{Engelmann1990}, CREAM \citep{Ahn2008}, \AMS-01 \citep{AMS01Nuclei2010} and \citet{Orth1978}. 
Fig.\,\ref{Fig::ccHeliumDeuteronXS} illustrates the uncertainty band derived by this procedure.
Note that this method has severe limitations and allows to simultaneusly explore only some parameters. 
More robust strategies require advanced statistical tools, see \textit{e.g.} \citet{Trotta2011}, \citet{Putze2010} 
and in particular \citet{Coste2011}. However the purpose in this work is estimating the parameter 
uncertainties rather than determining their exact values. 

The \textit{nuclear uncertainties} on the \Hd{} and \Het{} calculations are those arising from uncertainties 
in their production cross sections. In order to estimate the cross section uncertainties using the information 
from the measurements, we re-fit the parametrizations with the data to determine their overall normalizations
and associated errors. 
The uncertainty bands are shown in Fig.\,\ref{Fig::ccHeliumDeuteronXS} for the main reactions of \Hd{} and 
\Het{} production. These uncertainties can be directly translated into error bands for the predicted ratios 
\HdHeq{} and \HetHeq{}. These uncertainty bands are shown in Fig.\,\ref{Fig::ccIsotopeRatios2X2}.

The \AMS-01 data agree well with calculations within the astrophysical band, indicating consistency with the 
propagation picture arising from the \BC{} analysis. It is also clear that $Z\leq 2$ nuclear ratios carry valuable 
information on the transport parameters, \textit{i.e.}, they can be in principle used to tighten the 
constraints given by the \BC{} ratio. On the other hand, the \textit{nuclear uncertainties} represent
an intrinsic limitation on the accuracy of the model predictions, as reported in the right panels of the figure. 
Only precise cross section data or more refined calculations may pin down these uncertainties.
Unaccounted errors or systematic biases in cross section estimates cause errors on the 
predicted ratios which, in turn, may lead to a mis-determination of the CR transport parameters. 
Given the level of precision expected from PAMELA or \AMS-02, systematic errors in the cross section data 
may represent the dominant source of uncertainty for the model predictions of light CR isotopes.  
A strategy to check the model consistency with CR data is the use of secondary to secondary ratios such as \HdHet{}.
In fact, the \Hd{} and \Het{} isotopes have the same astrophysical origin and simlar progenitors (mainly \Heq{}). 
Thus, their ratio is almost unsensitive to the propagation physics and can be used to probe the net effect of the 
nuclear interactions. 
In fact, a mis-consistency between calculations and \HdHet{} data would indicate the presence of systematic biases 
in the cross sections that cannot be re-absorbed by the propagation parameters. 
As illustrated in the bottom panels of Fig.\,\ref{Fig::ccIsotopeRatios2X2}, 
the tight astrophysical uncertainty band (left) indicates a little discrepancy between data and model which can be understood 
if one consider the nuclear uncertainty (right), which is dominant for the \HdHet{} ratio. 
Similarly, the use of other ratios such as \LiRatio{}, \LiBe{} or \BRatio{} can represent a useful diagnostic tool 
for testing the overall consistency of the model.

%%%%%%%%%%%%%%%%%%%%%%%%%%%%%%
\section{Conclusions}      %%%
\label{Sec::Conclusions}   %%%
%%%%%%%%%%%%%%%%%%%%%%%%%%%%%%

We have compared new observations of the \HdHeq{} and \HetHeq{} ratios in CRs made by the 
\AMS-01 experiments with standard calculations of secondary production in the ISM. These ratios are well 
described by propagation models consistent with \BC{} ratio under a \textit{diffusive-reacceleration} 
scenario, suggesting the He and heavier nuclei such as \textsf{C-N-O} experience similar propagation histories.  
The accuracy of the secondary CR calculations relies on the accuracy of the cross sections employed.
Given the level of precision expected from \AMS-02, the errors in the cross section data 
will likely represent the dominant source of uncertainty for the model predictions of rare CR isotopes
such as \Hd{} or \Het{}. Similar issues may concern for $^{6,7}$\textsf{Li}, $^{7,9,10}$\textsf{Be}, or $^{10,11}$\textsf{B}.
These errors may be reduced using more refined calculations or more precise accelerator data.
For example, future observations may require the departure from the straight-ahead approximation 
which is generally assumed in the CR propagation studies for light nuclei.
Possible consistency checks for propagation models can be done using the secondary to secondary ratios, 
which are less sensitive to the astrophysical aspects of the interstellar CR propagation. The use of ratios 
such as \HdHet{}, \LiRatio{} or \BRatio{} can represent a useful diagnostic test for the 
reliability of the calculations: any CR propagation model, once tuned on secondary to primary ratios, must 
correctly reproduce the secondary to secondary ratios as well.
Another model limitation concerns the solar modulation effect.
Any more refined modeling requires to leave the force field approximation, that may be too simple to 
finely reproduce the future data of different $A/Z$ isotopes. 
Our understanding of the CR heliospheric propagation may be dramatically improved by the
\AMS-02 long-term observations of different CR species. 

%%%%%%%%%%%%%%%%%%%%%%%%%
 \section{Acknowledgments} %
%%%%%%%%%%%%%%%%%%%%%%%%%
I thank my many colleagues in \AMS{} for helpful discussions and
F. Focacci for her bibliographic support.
This work is supported by \textit{Agenzia Spaziale Italiana} under contract ASI-INFN 075/09/0. 

%%%%%%%%%%%%%%%%%%%%%%%%%%%%%%%

\end{document}